\documentclass[10pt,preprint]{aastex}

\shorttitle{The IR spectra of 2- and 9-vinylanthracene} 

\shortauthors{Maurya et al.}

\begin{document}

\title{Experimental and theoretical study on the infrared spectroscopy of \\
astrophysically relevant PAH derivatives 2- and 9-vinylanthracene}

\author{Anju Maurya, Shantanu Rastogi\altaffilmark{1}}
\affil{Department of Physics, Deen Dayal Upadhyay Gorakhpur University, Gorakhpur 273 009, India}
\email{shantanu\_r@hotmail.com}

\author{Ga\"{e}l Rouill\'e, Friedrich Huisken}
\affil{Laboratory Astrophysics Group of the Max Planck Institute for Astronomy at the Friedrich-Schiller-Universit\"at Jena, Institute of Solid State Physics, Helmholtzweg 3, D-07743 Jena, Germany}


\and

\author{Thomas Henning}
\affil{Max Planck Institute for Astronomy, K\"onigstuhl 17, D-69117 Heidelberg, Germany}

\altaffiltext{1}{Humboldt Fellow, Laboratory Astrophysics Group of the Max Planck Institute for Astronomy at the Friedrich-Schiller-Universit\"at Jena.}

\begin{abstract}

We propose to evaluate the contribution of polycyclic aromatic hydrocarbon molecules that carry side groups to the mid-infrared emission spectra. Within this framework, the IR absorption spectra of 2- and 9-vinylanthracene were measured in Ar matrices at 12 K and in CsI and polyethylene pellets at room temperature. The laboratory spectra were analyzed with the support of simulations based on the density functional theory. For each PAH molecule, eight IR spectra were computed by combining the B3LYP functional with as many different basis sets, namely 4-31G, 4-31G(d), 6-31G, 6-311G, 6-31G(d), 6-31G(d,p), 6-31+G(d,p), and 6-31++G(d,p). The comparison of the theoretical spectra with the laboratory data allowed us to determine the most suitable combinations for modeling the IR spectra of neutral PAH molecules that carry a vinyl side group. It was concluded from the examples of 2- and 9-VA that the optimum basis set is 6-31G unless a steric interaction has to be taken into account, in which case the optimum basis set is 6-31G(d). Thus, in the presence of such an interaction, the use of d-type polarization functions is recommended. We discuss the possibility for neutral vinyl-substituted PAHs to contribute to the mid-infrared emission spectra and find that their specific features do not match with the mid-infrared aromatic emission bands.

\end{abstract}

\keywords{ISM: lines and bands --- ISM: molecules --- molecular data}

\section{Introduction}

The astrophysical infrared emission features at 3.3, 6.2, 7.7, 8.6, 11.2, and 12.7 $\mu$m (3030, 1610, 1300, 1160, 890, and 790 cm$^{-1}$) are attributed to vibrational transitions in polycyclic aromatic hydrocarbon (PAH) molecules \citep{Leger84, Allamandola85, Allamandola89, Puget89}. These aromatic infrared bands (AIBs) are ubiquitously observed in a variety of locations ranging from star-forming regions to the vicinity of late type stars and even in external galaxies (\citealt{Cohen89, Geballe89, ISO96, Lutz98, Peeters02, Peeters04}; for a review, see \citealt{Tielens08}). The AIB profile variations correlate with object type indicating different PAH populations in different astrophysical regions \citep{Peeters02, Peeters04}. To understand the relationship between astrophysical conditions and types of PAHs, one needs to compare the astrophysical observations with reference data obtained in the laboratory. Studies of co-added infrared (IR) spectra of PAHs in different sizes and charge states provide useful indicators relating to possible PAH groups in the interstellar medium (ISM) \citep{Allamandola99, Pathak08}.

In the laboratory, the conditions of the ISM, which are low-temperature and collision-free environment, are best simulated with molecular beams. Because obtaining the IR emission spectra of PAHs under such conditions requires great efforts, laboratory data are scarce. Information on the vibrational modes of PAHs in molecular beams can be gained from the application of spectroscopy techniques that make use of IR absorption or dispersed fluorescence. The largest set of data, however, has been obtained for PAH molecules isolated in rare gas matrices \citep{Hudgins02}. Although the IR absorption spectra measured with the matrix isolation spectroscopy (MIS) technique are affected by the interaction between the molecules and the surrounding rare gas atoms, they provide useful data concerning the vibrational modes of PAHs. 

Laboratory studies of well identified PAH molecules are limited to those that can be synthesized and extracted at a reasonable cost. For the other ones, that is the majority of them, quantum theoretical calculations are used extensively to obtain absorption data \citep{Langhoff96, Bauschlicher00, Bauschlicher02, Pathak05, Pathak06, Pathak07, Bauschlicher08, Bauschlicher09}. The absorption data are used to model emission spectra and, by co-adding the emission bands of several PAHs, composite spectra can be obtained for a meaningful comparison with observations \citep{Allamandola99, Joblin02, Mulas06, Pathak08}.

Using quantum chemical data for a large set of PAHs, a good match for the complex 7.7 $\mu$m AIB from different sources was reported \citep{Pathak08}, pointing toward the existence of large PAHs in benign regions around planetary nebulae and small- to medium-sized PAHs in harsh UV-dominated, star forming regions. Most model spectra, however, failed to reproduce the 6.2 $\mu$m AIB as the closest calculated vibrations (CC stretching modes) had frequencies 30--40 cm$^{-1}$ too low \citep{Pathak08}. In the absence of simultaneous matching of all AIBs, it is essential to consider a wider set of PAHs, including derivatives, in emission models. Studies on substituted PAHs, hydrogenated ones, and PAHs with nitrogen heterocycles have been reported \citep{Langhoff98, Beegle01, Hudgins05}. Experimental studies on carbon nanoparticles produced by astrophysically relevant mechanisms suggest the presence in space of PAHs carrying side groups containing CC double bonds \citep{Hu06}. It is possible that such PAH derivatives are also formed in the ISM, and the presence of double bonds in the side group may increase the frequency of the relevant vibrations (in-plane deformations of the aromatic rings) and bring their bands closer to the 6.2 $\mu$m AIB. It is therefore important to study PAH vinyl derivatives, in order to incorporate them into AIB feature modelings.

Although quantum chemical methods have proven to be useful in providing IR data, they are restricted by the choice of basis sets and scaling procedures \citep{Martin96, Yoshida00}. It is usual to compare spectra of matrix-isolated samples of a few small PAHs with calculations using some basis set and use the same basis set and scaling factor for larger PAHs. The use of smaller basis sets, for instance 4-31G, is computationally less demanding and requires a simple linear scaling. Large systems can be efficiently studied with models based on the density functional theory (DFT). The B3LYP functional has been widely applied and numerous results obtained at the DFT-B3LYP/4-31G level of theory have been published \citep{Langhoff96, Bauschlicher00, Bauschlicher02, Pathak05, Pathak06, Pathak07, Bauschlicher08, Bauschlicher09}. These calculations give a good frequency match with experimental spectra, but the intensities, particularly those related to C$-$H stretching vibrations, are overestimated \citep{Langhoff96, Bauschlicher97, Hudgins98, Pathak05, Pathak06}. It has been shown that the use of larger basis sets provides a better intensity correlation but complicates the scaling procedure \citep{Bauschlicher97, Pathak06}. Because PAHs that have been studied to date were essentially plain ones, planar, with a single conformation, the DFT-B3LYP/4-31G model may not be the best to study PAHs with side groups.

In the present work, a combined experimental and theoretical study of the IR spectroscopy of 2-vinylanthracene (2-VA) and 9-vinylanthracene (9-VA) is reported. The structural diagram of the molecules is shown in Figure~\ref{fig}. The IR absorption spectra of both forms of VA isolated in Ar matrices are presented for the first time. Theoretical spectra have been calculated and are compared with the MIS measurements in order to evaluate the suitability of different basis sets and scaling laws. Finally the incorporation of vinyl-substituted PAHs into AIB models is discussed.

\section{Experimental}

Since the experimental setup was used in a recent study \citep{Rouille08}, only a short description is given here. The PAH-doped Ar matrices were formed on a cold KBr substrate placed in a vacuum chamber. A closed-cycle helium cryocooler allowed us to cool the substrate to temperatures as low as 7 K. Prior to forming each PAH-doped matrix, a reference spectrum of the windows and substrate was measured for the purpose of background correction.

The 2-VA (Chemos, purity 97\%) and 9-VA (Sigma-Aldrich, purity 97\%) samples were used as received. Since both 2-VA and 9-VA are solid under standard conditions, the sample under study was placed into a small oven connected by a short line to the vacuum chamber. Thus the vapor pressure of the sample could be increased by heating. When the oven reached the temperature chosen for the experiment, pure Ar gas (Linde, purity 99.9996\%) was injected in the line between the oven and the vacuum chamber. In excess, the rare gas was seeded with PAH molecules and its flow was directed toward the cold KBr substrate.

During the initial heating of the oven, the deposition of molecules was prevented by inserting a cold shield between the gas inlet and the cold KBr substrate. By the same means, and by stopping the carrier gas flow, it was also possible to interrupt the growth of the matrix at different times in order to measure spectra and to monitor the rise of absorption bands.

To produce each matrix, the temperature of the KBr substrate was stabilized at 12 K and the Ar flow was set to 3.70 sccm (standard cubic centimeter per minute). For 2-VA, the temperature of the oven had to be increased from 61.5 to 105.5~$\degr$C during a deposition process that took in total 105 min. In the case of 9-VA, the oven temperature was kept at 39.5~$\degr$C and the complete deposition lasted 90 min. The oven temperatures remained well below the melting points of the samples, 186.5--211~$\degr$C for 2-VA \citep{Hawkins57, Stolka76a} and 64--67~$\degr$C for 9-VA \citep{Hawkins57, Stolka76b}.

Each spectrum was measured in transmission through the matrix and the substrate with a Fourier transform IR spectrometer (Bruker 113v). Because mechanical vibrations caused by the helium compressor generated an intense noise at frequencies lower than 600 cm$^{-1}$, the spectra measured at the end of the deposition procedures were acquired after cooling the matrices down to 7 K and then switching off the compressor. During these final acquisitions, the temperature of the substrate increased from 7 to 20 K. All scans were carried out with a resolution of 2 cm$^{-1}$ and an averaging over 64 measurements.

The IR spectra of 2- and 9-VA were also measured in CsI and polyethylene pellets at room temperature. Note that in pellets the molecules are not isolated. They are found in sample grains embedded in the pellets. All pellets had a diameter of 13 mm and their masses were in the range 200--290 mg. The CsI and polyethylene pellets were prepared with mass ratios of 1:500 and 1:100, respectively. In polyethylene, measurements could be obtained down to 50 cm$^{-1}$.

\section{Results and discussion}

\subsection{Measured IR spectra}

The spectra of 2-VA and 9-VA measured at cryogenic temperatures in Ar matrices and at room temperature in CsI and polyethylene pellets are shown in Figure~\ref{fig1}. The strong bands observed in the spectrum of matrix-isolated 2-VA at 468, 739, and 885 cm$^{-1}$ remind us of the spectrum of anthracene \citep{Szczepanski93} while the strongest band at 900 cm$^{-1}$ is reminiscent of the main feature in the spectrum of ethylene \citep[H$_2$C$=$CH$_2$,][]{Cowieson81}. On the other hand, the features of anthracene are not obvious in the spectrum of 9-VA. For instance, the medium intensity band near 470 cm$^{-1}$ seems to be missing and the intensity of the band at 883 cm$^{-1}$ is reduced if compared with the intensity of the strongest feature at 735 cm$^{-1}$. Thus it appears that replacing a quarto hydrogen in anthracene with the vinyl group to form 2-VA does not affect the anthracene features, whereas the replacement of a central solo hydrogen to obtain 9-VA brings major modifications. Table~\ref{tbl-1} lists the frequencies of the bands measured with a relative intensity higher than 1\% of the strongest band.

Beside the most intense features, numerous weaker bands can be clearly seen in the matrix-isolated spectra of both 2- and 9-VA at frequencies up to 1550 cm$^{-1}$. The search for weak bands in the 1550--1700 cm$^{-1}$ range is hampered by the strong signal corresponding to the bending mode of matrix-isolated water molecules \citep{Ayers76}. Features corresponding to aromatic ring deformation modes and to the C=C stretching vibration of the vinyl group should be found in the 1600--1700 cm$^{-1}$ region. This is confirmed by the examination of the spectra measured in CsI pellets. They are free of water features and each shows near 1625 cm$^{-1}$ two bands of medium intensities, which overlap in the case of 9-VA. The examination of the water features at different times during the deposition of the PAH-doped matrices did not reveal superimposed PAH bands. Neither does the comparison of the water features observed in the spectra of 2- and 9-VA. One can conclude that the PAH bands expected in this region are well within the noise level and very weak. The column densities obtained with the present matrices would not allow us to observe these features even in the absence of the water lines.

At the high-frequency end of the spectra, one finds the characteristic peaks of the aromatic C$-$H stretching modes near 3050 cm$^{-1}$. As usually observed in PAH spectra, they have low intensities.

A study of jet-cooled 9-VA by dispersed emission spectroscopy reported low frequency modes at 90 and 134 cm$^{-1}$ \citep{Werst87}. We used polyethylene pellets containing 2- and 9-VA for measurements in the 50--220 cm$^{-1}$ frequency range. The spectra shown in Figure~\ref{fig1} do not reveal narrow bands in this interval. A weak, very broad feature may be present around 110 cm$^{-1}$ in the spectrum of 2-VA. A peak, although it is broad, appears more sharply at 124 cm$^{-1}$ in the spectrum of 9-VA.

\subsection{Theoretical calculations} \label{res:theo}

\begin{figure}
\epsscale{1.00}
\plotone{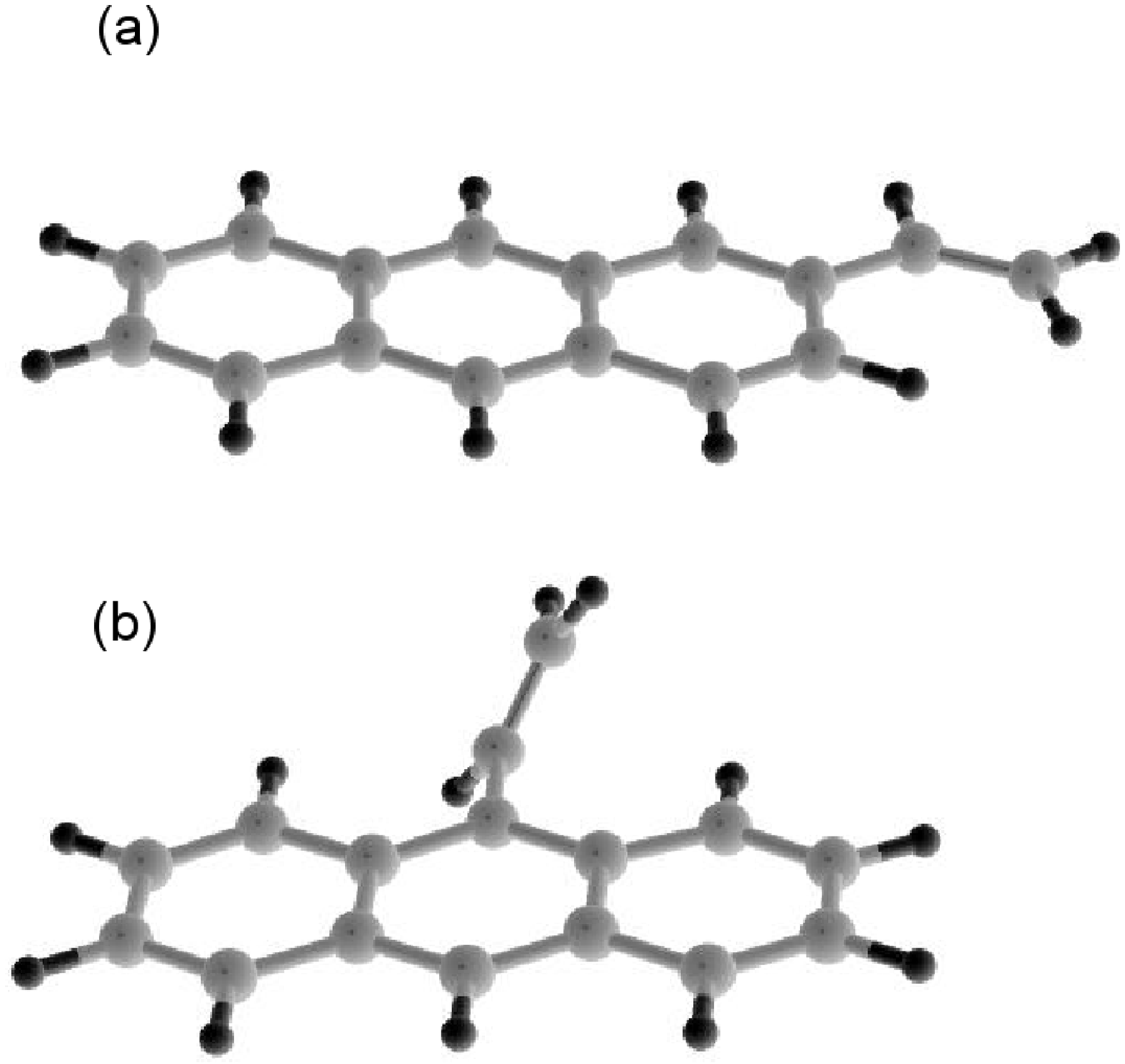}
\caption{Structural perspective of the two molecules (a) 2-vinylanthracene and (b) 9-vinylanthracene.\label{fig}}
\end{figure}

\begin{figure}
\epsscale{1.00}
\plotone{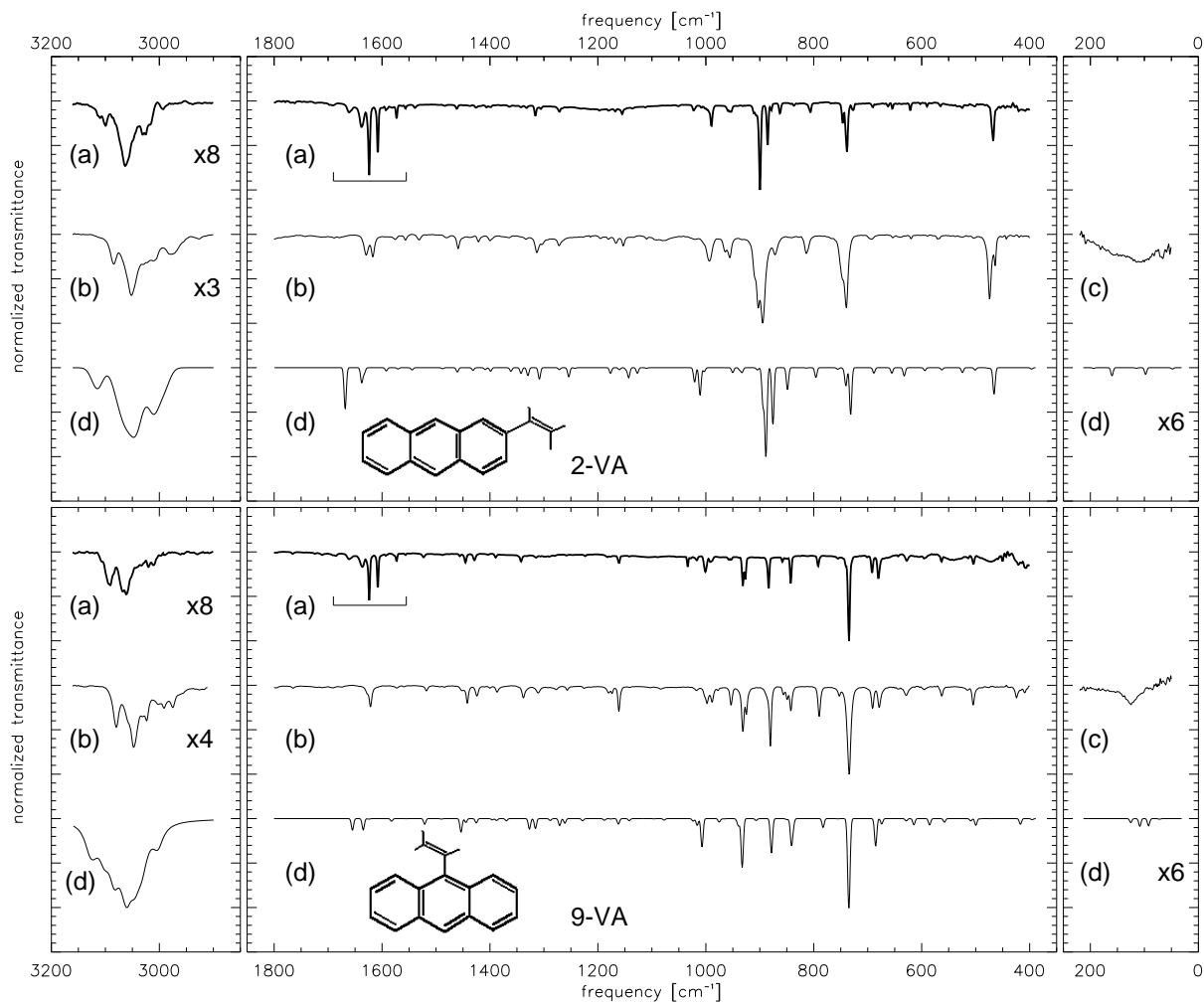}
\caption{Measured and theoretical absorption spectra of 2-VA (top) and 9-VA (bottom). The measured spectra were obtained under different conditions: molecules isolated in Ar matrices at 12 K (a), grains blended in CsI pellets at room temperature (b), and grains blended in polyethylene pellets at room temperature (c). The theoretical spectra (d) were computed at the B3LYP/6-31G(d,p) level and their frequency axes were scaled in a fitting procedure. In each MIS spectrum, a horizontal bracket marks the region of the H$_2$O features.\label{fig1}}
\end{figure}

Quantum chemical calculations have been performed with the GAMESS program \citep{Schmidt93}. The B3LYP functional has been employed in combination with eight different basis sets for both molecules to obtain optimized geometries and theoretical IR spectra. Figure~\ref{fig} shows the optimized structures. During the optimization procedures, all bond lengths and angles were let free. For 2-VA, however, successful optimization with the 6-31+G(d,p) basis set required a tightening of the integration grid. With the 6-31++G(d,p) basis set, it was necessary to confine the structure of 2-VA within a plane in order for its optimization to converge toward a solution. This constraint was consistent with the geometries obtained with the other basis sets.

We found that 2-VA had a planar geometry with the anti conformation if we use the convention chosen by \citet{Ni91b}. In our calculations, the energy of the syn conformation was determined to be approximately 400 cm$^{-1}$ higher. These results are in agreement with those obtained by \citet{Sakata01} with an ab initio approach at the Hartree-Fock level. Early experimental studies of the dynamics of the s-cis (syn) $\leftrightarrow$ s-trans (anti) isomerisation of 2-VA in solutions could not conclusively determine which conformer had the lowest energy \citep{Cherkasov62, Brearley85, Flom86, Arai89}. The outcome of our calculations, however, is the opposite of the theoretical and experimental results of \citet{Ni91a, Ni91b}. Using semiempirical models, with geometry constraints, and NMR measurements on deuterated 2-VA in solution, they concluded that a near planar syn conformation had the lowest energy \citep{Ni91a, Ni91b}. On the other hand, according to \citet{Cherkasov62}, the proportion of syn to anti isomers in a solution of 2-VA in heptane decreases with the temperature. For instance, it is 4.0 at 80~$\degr$C and 0.16 at $-$45~$\degr$C \citep{Cherkasov62}. In addition, NMR measurements gave conflicting results for 2-vinylnaphthalene, which was eventually found to have the lowest energy in the anti conformation \citep{Lewis96}. Relying on these results and on the fact that we considered a higher level of theory than \citet{Ni91a, Ni91b}, we accepted the anti conformation as the one with the lowest energy, which would be preponderant at cryogenic temperature. The relevance of our theoretical model was confirmed when comparing the spectra, calculated at both the anti and syn conformations, with the measured spectrum. All basis sets were tried in this procedure. Although the two conformations gave similar theoretical spectra, several bands of close intensities arose in the 400 -- 500 cm$^{-1}$ region for the syn conformation whereas a single band dominated the same region for the anti conformation. It was thus found that, in addition to the lowest theoretical energy, the anti conformation gave the theoretical spectra that agreed best with the measured spectrum. Accordingly, in order to interpret the spectrum of 2-VA isolated in an Ar matrix, the frequencies and IR activities of the vibrational modes for the anti conformation were retained. Table~\ref{tbl-2} lists the frequencies calculated using various basis sets and scaled after the spectrum was assigned as discussed in the next section. The frequency-scaled IR spectrum calculated with the 6-31G(d,p) basis set is displayed in Figure~\ref{fig1}.

In 9-VA, the optimized geometry for all used basis sets reveals planar moieties, but they make an angle of 56 $\pm$ 1$\degr$ due to steric interaction (Figure~\ref{fig}). This value is in excellent agreement with the experimental values 56.15 and 58.4$\degr$ obtained by X-ray diffraction measurements in crystals at 123 and 293 K, respectively \citep{Yasuda00}. It is also in good agreement with the angle of 60 $\pm$ 10$\degr$ reported in a study of the conformations of 9-substituted anthracenes, determined at room temperature in solution in carbon tetrachloride \citep{LeFevre68}. On the other hand, our result is not consistent with the angle of 0$\degr$ derived from dispersed fluorescence spectra of 9-VA measured in a supersonic jet \citep{Werst87}. We assume that the resolution of the jet-cooled spectrum is somehow responsible for this result. A previous computational study carried out with the AM1 and MNDO semiempirical models gave angles of 65$\degr$ and 90$\degr$ \citep{Ni91a}, respectively. Another study reporting the results of ab initio calculations at the Hartree-Fock level cited an angle of 69.9$\degr$ \citep{Sakata05}. The fact that the geometries we have calculated agree better than the others with the laboratory measurements of \citet{LeFevre68} and \citet{Yasuda00} can be attributed to the higher level of theory we have employed. Table~\ref{tbl-3} lists the scaled theoretical frequencies derived from our calculated geometries. As discussed in the next section, frequency scaling was carried out after the spectrum was assigned. The IR spectrum corresponding to the geometry obtained with the 6-31G(d,p) basis set is displayed in Figure~\ref{fig1}.

\begin{figure}
\epsscale{1.00}
\plotone{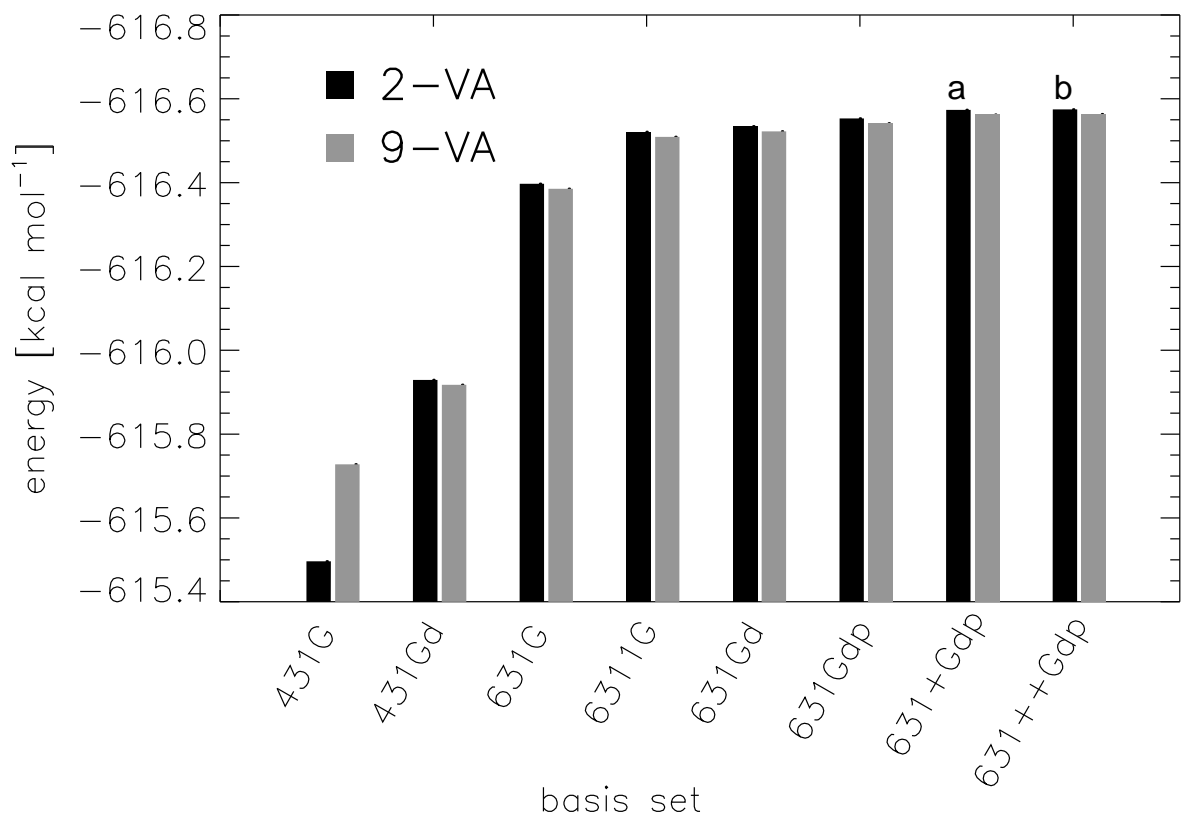}
\caption{Theoretical energies of 2- and 9-VA for geometries optimized with different basis sets. For 2-VA, a successful optimization required using a tight grid with the 631+Gdp basis set (a) and forcing a planar geometry with the 631++Gdp basis set (b).\label{fig2}}
\end{figure}

Figure~\ref{fig2} displays the theoretical energies of 2- and 9-VA as a function of the basis set used in each geometry optimization procedure. The figure shows expectedly that employing larger basis sets results in lower energies. Increasing the number of Gaussian primitives from four to six, e.g., by changing from the 4-31G basis set to 6-31G, gives the largest reduction in energy. The addition of d-type polarization functions on heavy atoms decreases the energy further as can be seen by comparing the energies obtained with the 4-31G and 4-31G(d) basis sets and those obtained with the 6-31G and 6-31G(d) basis sets. The introduction of p-type functions, however, does not diminish the energy significantly more if one considers the difference between the energies calculated using the 6-31G(d) and 6-31G(d,p) basis sets. Finally, adding diffuse functions to the description of C atoms or to both C and H atoms, which is indicated in the basis set notation with the $+$ and $++$ symbols, respectively, has also little effect on the theoretical energy. In the absence of experimental evidence, if one assumes that theoretical energies converge toward the physical value as they decrease, energy-wise there is little improvement beyond using the 6-31G(d) basis set. According to our calculations, the energies of 2- and 9-VA have similar values, the energy of 9-VA being slightly higher than that of 2-VA (except when using the 4-31G basis set), probably because of the supplementary steric interaction.

\subsection{Comparison between observed and calculated IR spectra} \label{res:comp}

The spectra of 2- and 9-VA isolated in Ar matrices were interpreted by comparing them with the raw theoretical spectra. Tables~\ref{tbl-2} and \ref{tbl-3} give the assignments for the bands observed in the spectra of 2- and 9-VA, respectively. The theoretical frequencies reported in Tables~\ref{tbl-2} and \ref{tbl-3} are the raw frequencies multiplied by a scaling factor. This factor depends on the theoretical model, that is on the basis set as far as this study is concerned. The determination of the scaling factor is discussed below.

In the case of 2-VA, the analysis of the observed spectrum confirms that the strongest band at 899.6 cm$^{-1}$ corresponds to a vibrational mode in which the main contribution is brought by the CH$_2$ wagging mode of the vinyl moiety. It is accompanied by the wagging of a pair of H atoms in the aromatic subunit. Concerning the strong bands at 468, 739, and 885 cm$^{-1}$, they arise from out-of-plane vibrations of the anthracene moiety as expected.

For 9-VA, the CH$_2$ wagging mode gives two bands of medium intensity at 926.6 and 931.5 cm$^{-1}$ due to its mixing with the anthracene CH wagging motions. This effect indicates a stronger interaction between the two moieties in 9-VA compared to 2-VA. The strongest band at 734.7 cm$^{-1}$ and the medium bands at 842.7 and 883.2 cm$^{-1}$ arise from out-of-plane vibrations in the anthracene subunit. Another pair of medium bands found at 680.8 and 693.3 cm$^{-1}$ corresponds to complex modes involving motions of all parts of 9-VA.

In the spectra of both 2- and 9-VA, the features associated with the aromatic C$-$H stretching modes were easily identified. The relative intensities of the observed bands is rather low whereas they are prominent features in the theoretical spectra obtained with the B3LYP functional and our selection of basis sets. This contrast was already observed for other PAHs \citep{Hudgins98,Bauschlicher97}.

The two bands observed near 1625 cm$^{-1}$ in the spectra of 2- and 9-VA measured in CsI pellets could be assigned. One band, which is found at 1617.0 and 1621.8 cm$^{-1}$ in the spectra of 2- and 9-VA, respectively, corresponds to an in-plane ring deformation mode, with a contribution of the C$=$C stretching in the case of 2-VA. It is related to the mode of anthracene measured at 1620 cm$^{-1}$ in a KBr pellet \citep{Karcher85} and at 1627 cm$^{-1}$ in an Ar matrix \citep{Szczepanski93}. The other band, found at 1629.6 cm$^{-1}$ for 2-VA and at 1627.6 cm$^{-1}$ for 9-VA, is attributed essentially to the C$=$C stretching vibration. Although the theoretical intensities obtained for these bands are rather strong, and despite the fact that they clearly appear in the spectra measured in CsI pellets, they are not found in the MIS measurements. Even though the water lines make the search difficult, only very weak bands can make it fruitless. In the spectra measured in CsI pellets, the intensities of these bands are assumed to be enhanced by the interaction of the molecules with their neighbors in the grains. For reference, the C=C stretching mode of ethylene in an Ar matrix has a frequency of 1629 cm$^{-1}$ \citep{Cowieson81}. Bands of medium strengths found near 1000 cm$^{-1}$ in the MIS spectra of 2- and 9-VA have been assigned to the torsion of the vinyl moiety about its double bond.

\begin{figure}
\epsscale{1.00}
\plotone{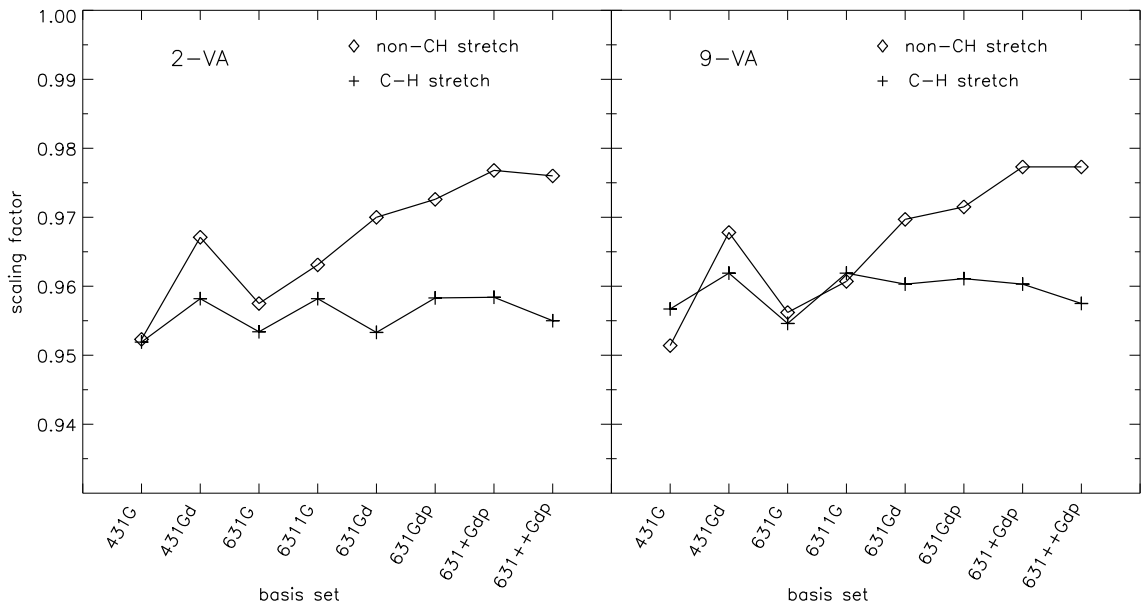}
\caption{Scaling factors obtained for theoretical spectra computed with different basis sets. The scaling factor for the CH stretching modes varies little with the basis set.\label{fig3}}
\end{figure}

After assigning the measured spectra, we determined the scaling factors to be applied to the theoretical frequencies. Following the example of studies on other PAHs \citep{Martin96, Bauschlicher97}, we considered separately the frequencies of the C$-$H stretching modes. These modes give rise to bands that strongly overlap resulting in the mostly unresolved 3.3 $\mu$m feature shown in Figure~\ref{fig1}. As it was not possible to assign 12 modes to this feature, frequency scaling factors were obtained by fitting a synthetic profile to the spectra measured in Ar matrix. For a given basis set, the synthetic profile was made of 12 Lorentzian functions, which were defined by the theoretical frequencies and intensities of the 12 C$-$H stretching vibrations. The Lorentzian functions were given a common full width at half maximum. Finally, they shared the same frequency and intensity scaling factors, which were determined in the fitting procedure. The complete assignment of measured frequencies to scaled theoretical harmonic frequencies is given in Tables~\ref{tbl-2} and \ref{tbl-3} for 2- and 9-VA, respectively. All scaling factors are given in Table~\ref{tbl-4}. They are also graphically presented as a function of the basis set in Figure~\ref{fig3}.

The description of the chemical bonds improves upon addition of polarization functions. This is reflected by the scaling factor for the 400--1800 cm$^{-1}$ range getting closer to 1 when a d-type polarization function is added to the 4-31G and 6-31G basis sets. For the aromatic C$-$H stretching frequencies, the scaling factor varies little as a function of the basis set. As a consequence, when using large basis sets, the frequency scaling required for the aromatic C$-$H stretching modes is more pronounced than that required for the other modes.

With the goal of determining which basis set to use so as to obtain the B3LYP-based theoretical IR spectra closest to observations, we calculated the deviations for the scaled frequencies and linear correlation coefficient ($R^2$) for relative intensities. Tables~\ref{tbl-5} and \ref{tbl-6} give for each molecule and for each basis set the root mean square (rms) value of the frequency differences, the maximum deviation observed for the frequencies, and the $R^2$ value for intensity correlation. This is reported for the 400--1800 cm$^{-1}$ range as the higher frequency C$-$H stretch region has weak peaks.

While the energy calculations show clear improvements on going to larger basis sets, with drastic change in going from four to six Gaussian functions (see Section~\ref{res:theo} and Figure~\ref{fig2}), there is no such clear indication in the frequency-intensity matching. For 2-VA, the best frequency match is obtained when using the tight integration grid in conjunction with the 6-31+G(d,p) basis set but the intensity is least correlated with the observed. Using the normal grid smallest frequency difference rms is obtained with 6-31G basis set while the intensity correlation initially did not appear to be good. On closer observation it is seen that with the 6-31G basis set there is a strong peak (911.6 cm$^{-1}$, relative intensity 0.96) calculated very close to the strongest one (905.6 cm$^{-1}$) and these are matched to the observed peaks at 911.2 (relative intensity 0.03) and 899.6 cm$^{-1}$ respectively. The 911.6 cm$^{-1}$ mode appears as an outlier in the intensity correlation. Disregarding this peak results in a good $R^2$ value which is mentioned in Table~\ref{tbl-5}.

In the case of 9-VA frequency matching improves with the basis set size but no such trend is there in intensity matching. In general for 9-VA the intensity matching is not as well correlated as for 2-VA. The best frequency-intensity match is obtained at the 6-31++G(d,p) level, the largest basis set used in the current study. For both 2- and 9-VA the intensity correlation is very good when 6-311G basis set is used but the corresponding frequency matching is poor.

\subsection{Astrophysical implications} \label{res:astro}

The detection of vinyl cyanide in \object{Sgr B2} provided the first evidence for the presence of the reactive vinyl radical in space \citep{Gardner75}. Ethylene itself was detected later in the circumstellar gas surrounding the supergiant \object{IRC+10216} \citep{Betz81}. Molecules containing the vinyl group have since been detected: vinyl alcohol \citep{Turner01} and propenal \citep{Hollis04} toward the star forming region \object{Sgr B2(N)}, and propylene (or propene) toward the dark cloud \object{TMC-1} \citep{Marcelino07}. As molecules containing a vinyl subunit have been discovered in different astrophysical environments, the study of 2- and 9-VA gains significance. 

We have expectedly found that the IR spectrum of a PAH is affected by the addition of a vinyl side group. Not only the vibrational modes of the vinyl moiety give rise to new bands, they also couple with the PAH modes. As a result, the IR bands of a vinyl-substituted PAH correlate with those of the non-substituted PAH through frequency shifts and intensity changes.

We have found that the CH$_2$ wagging mode gives rise to a band of strong intensity for 2-VA and medium intensity for 9-VA in both the theoretical and MIS spectra. Its wavelength, which is 11.1 $\mu$m (899.6 cm$^{-1}$) for 2-VA and $\sim$10.7 $\mu$m (931.4 cm$^{-1}$) for 9-VA, makes the band close to the 11.2 $\mu$m AIB. This region, however, is also rich in strong aromatic C--H wagging bands carried by non-substituted PAHs. As a consequence, the addition of the CH$_2$ wagging mode to AIB models may not improve them significantly. In that case, the existence of the 11.2 $\mu$m AIB would not be sufficient to prove the presence of vinyl-substituted PAHs in space. 

For both 2- and 9-VA, the DFT-based calculations have disclosed two bands of medium to strong intensity in the 6 $\mu$m region (see Figure~\ref{fig1}). One of them correlates with a band of anthracene, which is found at 6.15 $\mu$m in Ar matrix (1627 cm$^{-1}$, \citealt{Szczepanski93}) and at 6.17 $\mu$m in KBr pellet (1620 cm$^{-1}$, \citealt{Karcher85}). The other band involves the C$=$C stretching mode. The MIS measurements do not show these features, indicating that their intensity is significantly lower than predicted by the theoretical model we have chosen. Only the spectra measured in CsI pellets reveal bands in this region, possibly because their strength is enhanced by the interactions that take place between molecules in a grain. Caution is required to use these measurements. As can be seen in Table~\ref{tbl-1}, for a given band in the 1500 to 1600 cm$^{-1}$ interval (6.25 to 6.67 $\mu$m), the wavelength measured in a CsI pellet may be as much as 0.03 $\mu$m longer than the value obtained in an Ar matrix, which should be closer to the gas phase value. Consistently, the wavelength of the band of anthracene mentioned above is longer by 0.02 $\mu$m when measured on grains in a KBr pellet instead of molecules dispersed in an Ar matrix. As a consequence, it is for example difficult to evaluate the effect of the vinyl group addition on the mode that gives the band at 6.15 $\mu$m for anthracene isolated in an Ar matrix and at 6.17 and 6.18 $\mu$m (1621.8 and 1617.0 cm$^{-1}$) for 9- and 2-VA, respectively, in CsI pellets.

The two bands observed in CsI pellets in the 6 $\mu$m region lie close to the position of the 6.2 $\mu$m AIB. As the peak position of the 6.2 $\mu$m AIB varies depending on the line of sight, three classes of spectra were defined \citep{Peeters02}. Class A contains the 6.2 $\mu$m AIBs that peak at the shortest wavelengths, more specifically between 6.19 and 6.23 $\mu$m. In CsI pellets, the band correlated to anthracene is found at 6.17 and 6.18 $\mu$m (1621.8 and 1617.0 cm$^{-1}$) for 9- and 2-VA, respectively, just outside the short wavelength limit of the class A interval. The band connected to the C$=$C stretching mode is found at 6.136 and 6.144 $\mu$m (1629.6 and 1627.6 cm$^{-1}$) for 2- and 9-VA, respectively, well outside the limits of the class A interval. Obviously, it does not fit with the 6.2 $\mu$m AIB, even if considering the most favorable case of class A. To assume a redshift of 0.03 $\mu$m in the 6 $\mu$m region for the band positions measured in CsI pellets only increases the discrepancy.
 
Finally, the MIS spectra show bands close to 10 $\mu$m associated with the torsion of the vinyl moiety about its double bond. This mode is noticeably infrared active and should be seen in the presence of vinyl-substituted PAHs \citep{Papoular03}. An IR emission feature was observed at 10 $\mu$m but it was attributed to PAH cations without the mention of a particular structure \citep{Sloan99}.

The comparison of our laboratory spectra with the AIBs does not offer any evidence of the presence of vinyl-substituted PAHs in IR sources. Thus the population of these species is below the detection limit. Still, the vinyl radical has been included in schemes that model the growth of PAHs in carbon-rich stellar envelopes and outflows \citep{Frenklach89, Cherchneff92}. Let us assume that vinyl-substituted PAHs are formed in such environments and can be found in IR sources. In terms of size, 2- and 9-VA are small representatives of the PAH family. Theoretical studies, however, suggest that the AIBs are essentially caused by PAHs containing 80 C atoms or more \citep{Schutte93} and that only PAHs with more than 30 to 50 C atoms are not photodissociated by the UV field in the ISM \citep{Allain96a, Allain96b, Le Page03}. Consequently, the spectra of PAHs larger than 2- and 9-VA would be more suitable for a comparison with observed astrophysical spectra, especially if the action of a vinyl side group on the spectrum of a PAH molecule depends on the size of the latter.

\subsection{Computational aspects}

\begin{figure}
\epsscale{1.00}
\plotone{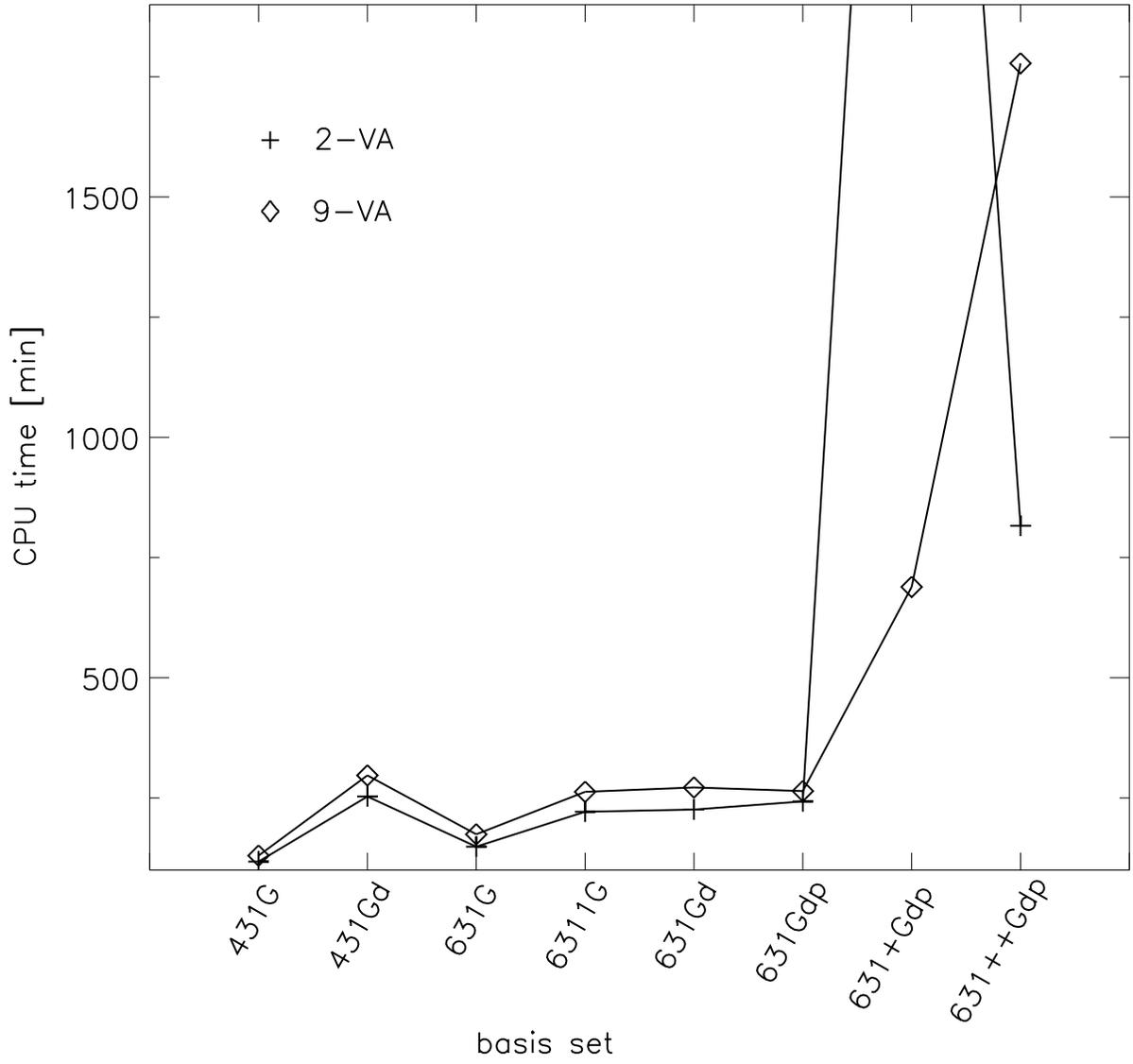}
\caption{CPU time required to calculate theoretical spectra with different basis sets. For 2-VA the tight grid 631+G(d,p) computation time 3994.2 min is out of range.\label{fig4}}
\end{figure}

To study larger vinyl-PAHs and PAHs with polyacetylenic side groups, which could be difficult to synthesize, it is important that a suitable level of quantum chemical theory be chosen. The study of both forms of VA and the comparison with the matrix-isolated IR spectra provide indicators for the correct choice. Ideally the use of ever larger basis sets tends to improve the accuracy of the calculations, but at the same time it drastically increases the computational effort. A compromise between accuracy and computational cost must be made. The duration of each vibrational calculation after geometry optimization is shown in Tables~\ref{tbl-5} and \ref{tbl-6}. Figure~\ref{fig4} proposes a graphical comparison. The clock times varied due to processor load and number of CPUs used. The data are extrapolated assuming eight CPUs (HP alpha 64 bit, 1.25 GHz) with 100\% usage. The computational time increases with the size of the basis set, especially upon addition of diffuse functions. In general, computations with a given basis set take less time for 2-VA (planar species) than for 9-VA. When using the 6-31+G(d,p) basis set to calculate the geometry of 2-VA, it was necessary to use the tight grid option for the determination of integrals. As the option was kept for calculating the vibrational modes, the computation time becomes excessively long. As the PAH size will increase, computations will be more time consuming.

The frequency-intensity matching in 2-VA shows that the 6-31G basis set is appropriate. For 9-VA also the 6-31G basis set gives reasonable result but the addition of d-type polarization function improves frequency matching and the intensity correlation also remains satisfactory. In 9-VA further improvements occur only on addition of diffuse functions but this will be computationally intensive. For both 2- and 9-VA calculations using d- or both d- and p-type polarization functions, the largest deviation from a measured frequency is observed for the $R($C$=$C$)_{\rm v}$ mode near 1600 cm$^{-1}$. In this region we used measurements obtained in CsI pellets that are likely to be redshifted compared with the value one would obtain by MIS in an Ar matrix (see Sections~\ref{res:comp} and \ref{res:astro}). This redshift could be 5--10 cm$^{-1}$ which is not enough to explain the overestimation in calculations for these modes. A different scaling factor in the 1600 cm$^{-1}$ region is indicated for calculations using basis sets with either d- or both d- and p-type polarization functions. There is greater overestimation of the $R($C$=$C$)_{\rm v}$ mode in 2-VA (planar species) than 9-VA (non-planar species). Considering all aspects the optimum basis set is 6-31G. When a steric interaction, as in 9-VA, has to be taken into account, the addition of d-polarization functions on C atoms can lead to improvements.

The frequency scaling factors for the same basis set are different for the two molecules reported. In the C$-$H stretch region the scaling factors 0.9534 (2-VA; 6-31G) and 0.9603 (9-VA; 6-31G(d)) can be used for the study of other similar species, with the same basis set. Similarly in the lower frequency range the scaling factors 0.9566 (2-VA; 6-31G) and 0.9697 (9-VA; 6-31G(d)) are suggested.

\section{Conclusions}

As few PAHs can be obtained and characterized in the laboratory, one has often to resort to theoretical approaches in order to evaluate IR spectra. Taking 2- and 9-VA as examples, we have examined the suitability of the DFT-B3LYP/4-31G level of theory to model the IR spectra of vinyl-substituted PAHs. At this level of theory, a good frequency match and a reasonable intensity match are obtained for non-substituted PAHs \citep{Langhoff96, Bauschlicher00, Bauschlicher02, Pathak05, Pathak06, Pathak07, Bauschlicher08, Bauschlicher09}. As the addition of an aliphatic side group may significantly alter the IR spectrum of a PAH, a more adapted level of theory may be required. We have found that, for an accurate description of the C$=$C bonds in a vinyl side group, the use of basis sets larger than 4-31G is indicated. The optimum basis set in our results is 6-31G for 2-VA and 6-31G(d) for 9-VA. Thus, in order to model the IR spectra of large vinyl-substituted PAHs, we propose to use the 6-31G basis set in the absence of any steric interaction, such as in 2-VA, and the 6-31G(d) basis set otherwise, such as in the case of 9-VA.

When the computational facilities allow it, one might be tempted to add diffuse functions to the basis sets to improve the theoretical spectra. Actually, the addition of diffuse functions on the C atoms and, better, on the C and H atoms, leads to improved theoretical spectra in the case of 9-VA. This could be interpreted as a better modeling of the molecule confronted with a steric interaction. Considering that a model that works in the presence of such an interaction should also work in its absence, this effect may be fortuitous. Indeed, for 2-VA, the addition of diffuse functions on the C atoms clearly lowers the quality of the theoretical band intensities. Diffuse functions need to be added to H atoms as well to find again a quality similar to that obtained without diffuse functions. If the improvement of the theoretical spectra lies on a better rendering of the steric interaction, it might be wiser to choose a different functional rather than to add diffuse functions to the basis sets. In comparison with the B3LYP functional, more recent functionals that include long range corrections, e.g. LC-wPBE \citep{Tawada04}, CAM-B3LYP \citep{Yanai04} and wB97X-D \citep{Chai-Head-Gordon08} need to be tried.

We have studied how the difference between an aromatic CC stretching mode and the 6.2 $\mu$m AIB is affected when taking into account PAHs carrying vinyl side groups, which contain a C$=$C bond. Because we could not observe bands around this wavelength in the spectra measured by MIS in Ar matrices, we rely on measurements obtained in CsI pellets and on the comparison with literature spectra of anthracene. It appears that the addition of a vinyl side group does not improve the match of the aromatic CC stretching mode with the 6.2 $\mu$m AIB. Moreover, the C$=$C stretching mode of the vinyl group gives rise to a new band in this region, which does not fit with the 6.2 $\mu$m AIB either.

The addition of vinyl groups introduces other new modes, notably the torsional mode about the C$=$C bond and the CH$_2$ wagging mode, which give bands near 10 and 11 $\mu$m, respectively. Astronomical observations, however, have not reported AIBs that could clearly correspond to these modes.

Further it needs to be studied whether the intensity of the bands induced by a vinyl side group increases in charged species, as in the case of aromatic C$-$C stretching modes in non-substituted PAHs \citep{Langhoff96, Bauschlicher00, Bauschlicher02, Pathak05, Pathak06, Pathak07}. In that case it would be possible to determine their correlation with AIBs.

\section{Acknowledgments}

The support of Alexander von Humboldt foundation to SR is appreciated. AM and SR acknowledge the financial assistance under ISRO-Respond project (ISRO/RES/2/336). The use of the High Performance Computing facility of the Inter-University Centre for Astronomy and Astrophysics, Pune, India is also acknowledged. The authors are grateful to Dr. H. Mutschke for providing the laboratory facilities and to G. Born for preparing the pellets and measuring their spectra.

\clearpage

\clearpage

\begin{deluxetable}{cccccccc}
\tabletypesize{\scriptsize}
\tablecaption{Vibrational Bands of 2- and 9-VA\label{tbl-1}}
\tablewidth{0pt}
\tablehead{
\multicolumn{4}{c}{2-VA} & \multicolumn{4}{c}{9-VA} \\
\multicolumn{2}{c}{Ar Matrix} & \multicolumn{2}{c}{CsI Pellet} & \multicolumn{2}{c}{Ar Matrix} & \multicolumn{2}{c}{CsI Pellet} \\
\colhead{Frequency} & \colhead{Intensity} & \colhead{Frequency} & \colhead{Intensity} & \colhead{Frequency} & \colhead{Intensity} & \colhead{Frequency} & \colhead{Intensity} \\
\colhead{(cm$^{-1}$)} &  & \colhead{(cm$^{-1}$)} &  & \colhead{(cm$^{-1}$)} &  & \colhead{(cm$^{-1}$)} & 
}
\startdata
3109.7 & 0.025 &        &       &     3091.3 & 0.050 & 3079.8 & 0.073 \\
3100.0 & 0.017 & 3084.6 & 0.109 &     3068.2 & 0.059 &        &       \\
3063.4 & 0.085 & 3051.8 & 0.241 &     3061.4 & 0.062 & 3047.0 & 0.130 \\
3030.6 & 0.033 &        &       &            &       & 3031.6 & 0.044 \\
3025.8 & 0.032 &        &       &     3019.0 & 0.010 & 3023.8 & 0.055 \\
3016.1 & 0.017 & 3011.3 & 0.111 &     3011.0 & 0.013 &        &       \\
       &       & 2976.6 & 0.092 &            &       & 1627.6 & 0.076 \\ 
       &       & 1629.6 & 0.207 &            &       & 1621.8 & 0.221 \\
       &       & 1617.0 & 0.230 &            &       & 1572.7 & 0.024 \\
1581.3 & 0.008 & 1575.6 & 0.033 &            &       & 1559.2 & 0.007 \\
1563.0 & 0.006 & 1556.3 & 0.048 &     1523.5 & 0.039 & 1517.7 & 0.044 \\
1538.9 & 0.032 & 1531.2 & 0.053 &     1487.8 & 0.008 & 1484.0 & 0.009 \\ 
1483.0 & 0.012 & 1480.1 & 0.025 &     1456.0 & 0.017 & 1452.1 & 0.027 \\
1461.8 & 0.045 & 1458.9 & 0.126 &     1445.4 & 0.086 & 1441.5 & 0.184 \\
1437.7 & 0.001 & 1437.7 & 0.011 &     1429.0 & 0.076 & 1424.2 & 0.103 \\
1426.1 & 0.023 & 1421.3 & 0.053 &     1403.9 & 0.009 & 1400.1 & 0.015 \\
1406.8 & 0.028 &        &       &     1389.5 & 0.036 & 1386.6 & 0.059 \\
1400.1 & 0.017 & 1400.1 & 0.040 &     1342.2 & 0.065 & 1338.4 & 0.124 \\
1364.4 & 0.015 & 1364.4 & 0.020 &     1315.2 & 0.018 & 1310.4 & 0.069 \\
1343.2 & 0.014 &        &       &     1280.5 & 0.006 & 1277.6 & 0.024 \\
1332.6 & 0.020 & 1333.5 & 0.021 &     1258.3 & 0.007 & 1256.4 & 0.035 \\ 
1316.2 & 0.123 & 1313.3 & 0.157 &     1223.6 & 0.018 & 1225.5 & 0.017 \\
1305.6 & 0.027 & 1303.6 & 0.025 &     1183.1 & 0.023 & 1180.2 & 0.059 \\
1270.9 & 0.056 & 1272.8 & 0.070 &     1177.3 & 0.009 & 1173.5 & 0.077 \\
1197.6 & 0.011 & 1195.6 & 0.019 &     1160.9 & 0.065 & 1160.9 & 0.191 \\
1180.2 & 0.017 & 1181.2 & 0.022 &     1142.6 & 0.005 & 1141.7 & 0.007 \\
1167.7 & 0.026 & 1166.7 & 0.040 &     1089.6 & 0.001 & 1083.8 & 0.015 \\ 
1155.2 & 0.060 & 1153.2 & 0.082 &     1028.8 & 0.007 & 1028.8 & 0.003 \\
1135.9 & 0.003 & 1133.9 & 0.008 &     1017.3 & 0.037 & 1017.3 & 0.031 \\
1112.7 & 0.002 & 1109.8 & 0.024 &     1000.9 & 0.126 &  997.9 & 0.192 \\
1023.1 & 0.062 & 1020.2 & 0.013 &      992.2 & 0.059 &  988.3 & 0.180 \\
1008.6 & 0.016 &        &       &      957.5 & 0.031 &        &       \\
 989.3 & 0.233 &  993.2 & 0.252 &      953.6 & 0.023 &  953.6 & 0.211 \\
 958.4 & 0.063 &  964.2 & 0.153 &      931.4 & 0.369 &  931.4 & 0.515 \\
 954.6 & 0.068 &  955.6 & 0.214 &      926.6 & 0.231 &  924.7 & 0.319 \\
 911.2 & 0.030 &        &       &      883.2 & 0.390 &  880.3 & 0.680 \\
 899.6 & 1.000 &  902.5 & 0.818 &      858.2 & 0.041 &  856.2 & 0.050 \\
 885.2 & 0.459 &  894.8 & 1.000 &      842.7 & 0.359 &  842.7 & 0.280 \\
 863.0 & 0.114 &  871.7 & 0.180 &      791.6 & 0.107 &  789.7 & 0.344 \\
 837.0 & 0.019 &        &       &      755.0 & 0.014 &  753.1 & 0.049 \\
 806.1 & 0.099 &  813.8 & 0.183 &      740.5 & 0.044 &        &       \\
 768.5 & 0.010 &        &       &      734.7 & 1.000 &  734.7 & 1.000 \\                                   
 746.3 & 0.250 &  746.3 & 0.076 &      692.3 & 0.183 &  691.4 & 0.173 \\
 738.6 & 0.585 &  739.6 & 0.848 &      680.8 & 0.230 &  678.8 & 0.246 \\
 690.4 & 0.037 &  693.3 & 0.036 &      644.1 & 0.015 &  641.2 & 0.009 \\  
 654.7 & 0.067 &  653.8 & 0.025 &      627.7 & 0.065 &  628.7 & 0.107 \\
 621.0 & 0.089 &  620.0 & 0.024 &      595.9 & 0.031 &  594.9 & 0.033 \\
 590.1 & 0.042 &  589.1 & 0.010 &      563.1 & 0.057 &  563.1 & 0.114 \\
 565.0 & 0.035 &  570.8 & 0.034 &      513.9 & 0.020 &  514.9 & 0.026 \\
 525.5 & 0.028 &        &       &      504.3 & 0.090 &  504.3 & 0.218 \\
 501.4 & 0.025 &  503.3 & 0.021 &      420.4 & 0.069 &  424.3 & 0.033 \\
 467.7 & 0.427 &  474.4 & 0.714 &      406.9 & 0.050 &  405.9 & 0.095 \\     
\enddata
\tablecomments{The intensities are normalized with respect to the strongest band of each set.}
\end{deluxetable}

\clearpage

\begin{deluxetable}{cccccccclcc}
\tabletypesize{\scriptsize}
\rotate
\tablecaption{Assignment of the Vibrational Bands of 2-VA\label{tbl-2}}
\tablewidth{0pt}
\tablehead{
\multicolumn{8}{c}{Theoretical Frequency\tablenotemark{a}} & \colhead{Mode Description\tablenotemark{b}} & \colhead{Mode No.\tablenotemark{b}} & \colhead{Observed Frequency\tablenotemark{c}} \\
\multicolumn{8}{c}{(cm$^{-1}$)} &  &  & \colhead{(cm$^{-1}$ / ($\mu$m))} \\
\colhead{431G} & \colhead{431Gd} & \colhead{631G} & \colhead{6311G} & \colhead{631Gd} & \colhead{631Gdp} & \colhead{631+Gdp\tablenotemark{d}} & \colhead{631++Gdp\tablenotemark{e}} &  &  & 
}
\startdata
3095.2 & 3108.1 & 3118.6 & 3087.8 & 3101.6 & 3124.4 & 3111.1 & 3102.2 & $r_{\rm a}({\rm CH}_2)_{\rm v}$                               & 78 & 3109.7 (3.22)\\
3071.7 & 3079.0 & 3086.5 & 3072.6 & 3072.3 & 3087.5 & 3077.3 & 3073.6 & $r({\rm CH})$                                                 & 77 & 3100.0 (3.23) \\
3058.6 & 3067.8 & 3070.7 & 3059.5 & 3060.4 & 3072.8 & 3070.6 & 3062.4 & $r({\rm CH})$                                                 & 76 & 3063.4 (3.26) \\
3053.0 & 3054.8 & 3057.4 & 3053.7 & 3055.7 & 3062.4 & 3057.4 & 3057.6 & $r({\rm CH})$                                                 & 75 &        \\
3045.6 & 3052.1 & 3052.0 & 3045.2 & 3049.6 & 3054.9 & 3053.7 & 3051.3 & $r({\rm CH})$                                                 & 74 &        \\
3040.4 & 3048.8 & 3050.6 & 3040.6 & 3044.8 & 3053.3 & 3052.5 & 3046.7 & $r({\rm CH})$                                                 & 73 &        \\
3016.7 & 3026.0 & 3040.8 & 3007.9 & 3024.1 & 3046.0 & 3020.7 & 3019.2 & $r({\rm CH}_2)_{\rm v}$                                       & 72 &      \\
3037.0 & 3040.9 & 3036.5 & 3037.2 & 3041.0 & 3040.3 & 3040.8 & 3042.7 & $r({\rm CH})$                                                 & 71 & 3030.6 (3.30) \\
3008.5 & 3024.8 & 3019.0 & 3009.7 & 3012.6 & 3023.6 & 3026.0 & 3014.9 & $r({\rm CH})$                                                 & 70 & 3025.8 (3.30) \\
3001.3 & 3021.5 & 3013.7 & 3002.3 & 3005.6 & 3018.1 & 3023.9 & 3007.3 & $r({\rm CH})$                                                 & 69 & 3016.1 (3.32)\\
2996.3 & 3013.0 & 3009.0 & 2997.1 & 3000.6 & 3013.5 & 3017.0 & 3002.5 & $r({\rm CH})$                                                 & 68 &        \\
2974.3 & 3034.5 & 2988.8 & 2976.0 & 2984.6 & 2999.7 & 3034.9 & 2987.8 & $r({\rm CH})_{\rm v}$                                         & 67 & 2976.6\tablenotemark{f} (3.36) \\
1643.0 & 1670.4 & 1644.3 & 1634.3 & 1666.8 & 1668.2 & 1650.8 & 1657.3 & $R($C$=$C$)_{\rm v}+\alpha({\rm HCH})_{\rm v}+\beta({\rm CCH})_{\rm v}$    & 66 & 1629.6\tablenotemark{f} (6.14) \\
1610.8 & 1636.0 & 1614.8 & 1609.5 & 1636.3 & 1637.6 & 1632.9 & 1633.7 & $R({\rm CC}) + \beta({\rm CCH})$ & 65 & 1617.0\tablenotemark{f} (6.18) \\
1604.7 & 1629.5 & 1612.0 & 1606.6 & 1630.1 & 1632.7 & 1630.6 & 1630.5 & $R({\rm CC}) + \beta({\rm CCH})$ & 64 &        \\
1568.7 & 1591.2 & 1571.6 & 1566.4 & 1590.4 & 1592.0 & 1589.2 & 1588.3 & $R({\rm CC}) + \beta({\rm CCH})$ & 63 & 1581.3 (6.32) \\
1540.3 & 1566.9 & 1547.2 & 1542.6 & 1568.6 & 1570.5 & 1561.3 & 1569.3 & $R({\rm CC}) + \beta({\rm CCH})$ & 62 & 1563.0 (6.40) \\
1520.5 & 1542.9 & 1525.6 & 1520.4 & 1543.2 & 1544.2 & 1541.2 & 1541.9 & $R({\rm CC}) + \beta({\rm CCH}) + \alpha({\rm HCH})_{\rm v}$  & 61 & 1538.9 (6.50) \\
1474.8 & 1488.8 & 1477.2 & 1476.3 & 1489.1 & 1487.6 & 1487.1 & 1485.7 & $R({\rm CC}) + \beta({\rm CCH})$ & 60 & 1483.0 (6.74)\\
1456.3 & 1463.3 & 1455.8 & 1456.9 & 1462.8 & 1460.3 & 1462.6 & 1458.7 & $R({\rm CC}) + \beta({\rm CCH}) + \alpha({\rm HCH})_{\rm v} + \beta({\rm CCH})_{\rm v}$ & 59 & 1461.8 (6.84) \\
1443.8 & 1448.3 & 1442.5 & 1442.8 & 1447.7 & 1444.5 & 1442.0 & 1442.7 & $R({\rm CC}) + \beta({\rm CCH}) + \alpha({\rm HCH})_{\rm v} + \beta({\rm CCH})_{\rm v}$ & 58 & 1437.7 (6.96) \\
1423.6 & 1433.2 & 1425.5 & 1424.2 & 1433.7 & 1431.4 & 1425.5 & 1429.4 & $R({\rm CC}) + \beta({\rm CCH}) + \alpha({\rm HCH})_{\rm v} + \beta({\rm CCH})_{\rm v}$ & 57 & 1426.1 (7.01) \\
1391.2 & 1406.6 & 1399.5 & 1391.1 & 1408.7 & 1408.8 & 1403.2 & 1408.1 & $R({\rm CC}) + \alpha({\rm HCH})_{\rm v} + \beta({\rm CCH})_{\rm v}$ & 56 & 1406.8 (7.11)\\
1381.4 & 1395.0 & 1390.6 & 1380.8 & 1398.4 & 1398.8 & 1391.8 & 1399.4 & $R({\rm CC}) + \beta({\rm CCH})$ & 55 & 1400.1 (7.14) \\
1363.2 & 1364.7 & 1363.0 & 1361.5 & 1363.1 & 1361.1 & 1367.6 & 1360.6 & $R({\rm CC}) + \beta({\rm CCH}) + \beta({\rm CCH})_{\rm v}$ & 54 & 1364.4 (7.33) \\
1338.2 & 1340.9 & 1339.2 & 1338.1 & 1342.7 & 1342.4 & 1344.8 & 1342.4 & $R({\rm CC}) + \beta({\rm CCH}) + \alpha({\rm HCH})_{\rm v} + \beta({\rm CCH})_{\rm v}$ & 53 & 1343.2 (7.44) \\
1325.4 & 1333.9 & 1328.8 & 1325.0 & 1332.4 & 1329.8 & 1310.7 & 1329.6 & $R({\rm CC}) + \beta({\rm CCH}) + \beta({\rm CCH})_{\rm v}$ & 52 & 1332.6 (7.50) \\
1302.9 & 1310.3 & 1303.7 & 1302.2 & 1309.0 & 1308.5 & 1301.9 & 1308.2 & $R({\rm CC}) + \beta({\rm CCH}) + \beta({\rm CCH})_{\rm v}$ & 51 & 1316.2 (7.60) \\
1263.4 & 1271.4 & 1267.0 & 1259.7 & 1271.2 & 1271.4 & 1271.7 & 1269.4 & $R({\rm CC}) + R({\rm CC})_{\rm v} + \beta({\rm CCH}) + \alpha({\rm CCC})$  & 50 & 1305.6 (7.66) \\
1266.3 & 1258.7 & 1264.1 & 1264.3 & 1255.7 & 1254.0 & 1261.8 & 1255.5 & $R({\rm CC}) + \beta({\rm CCH}) + \alpha({\rm CCC})$  & 49 & 1270.9 (7.87) \\
1179.3 & 1179.4 & 1179.4 & 1177.2 & 1178.0 & 1176.6 & 1195.3 & 1176.8 & $R({\rm CC})_{\rm v} + \beta({\rm CCH}) + \beta({\rm CCH})_{\rm v}$  & 47 & 1197.6 (8.35)\\
1168.3 & 1164.6 & 1170.0 & 1167.3 & 1163.3 & 1160.2 & 1180.4 & 1160.2 & $\beta({\rm CCH})$ & 46 & 1180.2 (8.47) \\
1154.9 & 1150.7 & 1155.5 & 1153.3 & 1147.1 & 1147.5 & 1166.5 & 1145.6 & $\beta({\rm CCH})$ & 45 & 1167.7 (8.56) \\
1148.1 & 1146.3 & 1149.2 & 1146.1 & 1146.3 & 1142.8 & 1153.5 & 1143.7 & $\beta({\rm CCH})$ & 44 & 1155.2 (8.66) \\
1140.5 & 1129.4 & 1140.9 & 1133.8 & 1129.8 & 1126.9 & 1144.1 & 1127.2 & $\beta({\rm CCH})$ & 43 & 1135.9 (8.80)\\
1104.0 & 1111.1 & 1105.4 & 1102.3 & 1108.9 & 1110.0 & 1111.1 & 1108.1 & $\beta({\rm CCH}) + \alpha({\rm CCC})$ & 42 & 1112.7 (8.99)\\
1030.7 & 1021.4 & 1025.5 & 1032.4 & 1027.4 & 1020.7 & 1014.1 & 1025.8 & $\beta({\rm HCH})_{\rm v} + \beta({\rm CCH})_{\rm v}$ & 41 & 1023.1 (9.77) \\
1010.0 & 1010.6 & 1010.1 & 1014.6 & 1010.7 & 1010.7 & 999.6  & 1005.7 & $R({\rm CC}) + \beta({\rm CCH})$ & 40 & 989.3 (10.11)\\
994.4  & 1002.8 & 995.4  & 991.7  & 1003.8 & 1003.0 & 1007.6 & 1003.3 & $\epsilon({\rm CH})_{\rm v} + \omega({\rm HCH})_{\rm v}$ & 39 & 1008.6 (9.91) \\
953.2  & 949.9  & 950.7  & 953.0  & 950.2  & 950.2  & 944.6  & 952.6  & $\beta({\rm CCH}) + \alpha({\rm CCC})$ & 37 & 958.4 (10.43)\\
956.5  & 936.1  & 952.6  & 959.6  & 931.7  & 937.0  & 956.4  & 950.1  & $\epsilon({\rm CH})$ & 36 & 954.6 (10.48)\\
948.1  & 930.7  & 944.4  & 954.1  & 926.0  & 932.8  & 948.8  & 943.5  & $\epsilon({\rm CH})$ & 35 &       \\
913.9  & 906.8  & 911.6  & 913.5  & 901.0  & 904.4  & 898.4  & 904.4  & $\alpha({\rm CCC})$ & 34 & 911.2 (10.97)\\
916.5  & 890.3  & 915.5  & 923.0  & 891.1  & 894.1  & 902.2  & 897.6  & $\epsilon({\rm CH}) + \epsilon({\rm HCH})_{\rm v}$ & 33 &     \\
903.7  & 884.9  & 905.6  & 903.9  & 886.5  & 888.6  & 900.3  & 893.8  & $\epsilon({\rm CH}) + \epsilon({\rm HCH})_{\rm v}$ & 32 & 899.6 (11.12)\\
887.1  & 871.5  & 888.9  & 888.9  & 874.6  & 875.7  & 885.9  & 878.6  & $\epsilon({\rm CH})$ & 31 & 885.2 (11.30)\\
856.8  & 844.6  & 857.6  & 860.4  & 845.9  & 848.7  & 860.1  & 852.5  & $\epsilon({\rm CH})$ & 30 & 863.0 (11.59)\\
807.0  & 813.2  & 807.5  & 806.4  & 811.9  & 813.3  & 821.6  & 814.3  & $R({\rm CC}) + \alpha({\rm CCC})$ & 28 & 837.0 (11.95)\\
799.3  & 792.1  & 801.2  & 802.5  & 794.8  & 796.1  & 802.0  & 795.5  & $\epsilon({\rm CH})$ & 27 & 806.1 (12.41)\\
757.4  & 753.2  & 756.3  & 750.3  & 754.3  & 754.8  & 753.8  & 752.8  & $\epsilon({\rm CH})$ & 25 & 768.5 (13.01)\\
742.6  & 743.7  & 741.9  & 742.4\tablenotemark{g}  & 743.5  & 744.3  & 741.2  & 745.5  & $\epsilon({\rm CH})$ & 24 &       \\
742.3  & 737.5  & 742.4  & 742.3  & 739.2  & 740.5  & 740.6  & 738.7  & $\epsilon({\rm CH}) + \tau({\rm CCC})$ & 23 & 746.3 (13.40)\\
731.9  & 727.7  & 733.1  & 731.9  & 729.9  & 731.6  & 732.0  & 729.6  & $\epsilon({\rm CH}) + \tau({\rm CCC})$ & 22 & 738.6 (13.54)\\
686.5  & 684.8  & 688.5  & 689.6  & 687.4  & 688.6  & 689.0  & 686.6  & $\epsilon({\rm CH}) + \tau({\rm CCC}) + \omega({\rm HCH})_{\rm v}$ & 21 & 690.4 (14.48)\\
662.9  & 656.2  & 659.9  & 664.0  & 654.8  & 655.4  & 648.5  & 657.1  & $\alpha({\rm CCC}) + \beta({\rm HCH})_{\rm v}$ & 20 & 654.7 (15.27)\\
641.3  & 634.0  & 638.2  & 642.9  & 631.6  & 632.4  & 619.8  & 633.5  & $\alpha({\rm CCC})$ & 19 & 621.0 (16.10)\\
601.4  & 593.2  & 599.5  & 602.5  & 592.8  & 594.2  & 586.0  & 594.4  & $\alpha({\rm CCC}) + \alpha({\rm CCC})_{\rm v} + \beta({\rm HCH})_{\rm v}$ & 18 & 590.1 (16.95) \\
567.2  & 561.8  & 565.1  & 568.3  & 561.3  & 562.6  & 562.7  & 563.2  & $\tau({\rm CCC}) + \omega({\rm HCH})_{\rm v}$ & 17 & 565.0 (17.70)\\
532.0  & 523.9  & 529.6  & 532.4  & 523.4  & 524.6  & 519.3  & 525.2  & $\alpha({\rm CCC}) + \alpha({\rm CCC})_{\rm v} + \beta({\rm HCH})_{\rm v}$ & 16 & 525.5 (19.03)\\
501.2  & 498.0  & 502.3  & 505.0  & 500.0  & 501.0  & 501.1  & 502.2  & $\tau({\rm CCC}) + \omega({\rm HCH})_{\rm v}$ & 15 & 501.4 (19.94)\\
470.6  & 469.9  & 472.0  & 474.3  & 471.7  & 472.6  & 473.9  & 472.7  & $\tau({\rm CCC})$ & 13 &       \\
462.6  & 463.0  & 464.8  & 467.1  & 465.6  & 465.9  & 467.7  & 464.3  & $\tau({\rm CCC})$ & 12 & 467.7 (21.38)\\
\enddata
\tablenotetext{a}{The frequencies have been scaled.}
\tablenotetext{b}{Derived from the results obtained with the 631Gdp basis set. Notation of modes according to \citet{Banisaukas04}:
$r({\rm CH})$, aromatic CH stretch;
$r({\rm CH})_{\rm v}$, vinyl CH stretch;
$r({\rm CH}_2)_{\rm v}$, symmetric vinyl CH$_2$ stretch;
$r_{\rm a}({\rm CH}_2)_{\rm v}$, asymmetric vinyl CH$_2$ stretch;
$R({\rm CC})$, aromatic CC stretch;
$R({\rm CC})_{\rm v}$, vinyl-ring junction CC stretch;
$R({\rm C}={\rm C})_{\rm v}$, vinyl C$=$C stretch;
$\alpha({\rm CCC})$, in-plane aromatic ring angular deformation;
$\alpha({\rm CCC})_{\rm v}$, in-plane vinyl CCC bend;
$\beta({\rm CCH})$, in-plane aromatic CH bend;
$\beta({\rm CCH})_{\rm v}$, in-plane vinyl CH bend;
$\epsilon({\rm CH})$, out-of-plane aromatic CH bend;
$\epsilon({\rm CH})_{\rm v}$, out-of-plane vinyl CH bend;
$\alpha({\rm HCH})_{\rm v}$, vinyl HCH scissor;
$\beta({\rm HCH})_{\rm v}$, vinyl HCH rock;
$\epsilon({\rm HCH})_{\rm v}$, vinyl HCH out-of-plane wag;
$\omega({\rm HCH})_{\rm v}$, vinyl HCH out-of-plane twist;
$\tau({\rm CCC})$, out-of-plane aromatic ring angular deformation;
$\tau({\rm CCC})_{\rm v}$, out-of-plane angular deformation in the vinyl moiety.}
\tablenotetext{c}{Frequencies measured in Ar matrix unless indicated otherwise.}
\tablenotetext{d}{Geometry optimized using a tight grid.}
\tablenotetext{e}{Geometry optimized forcing a planar geometry.}
\tablenotetext{f}{Band observed at room temperature in CsI pellet.}
\tablenotetext{g}{The atomic displacements ($\epsilon({\rm CH})$) differ from those obtained with the other basis sets.}
\end{deluxetable}

\clearpage

\begin{deluxetable}{cccccccclcc}
\tabletypesize{\scriptsize} 
\rotate
\tablecaption{Assignment of the Vibrational Bands of 9-VA\label{tbl-3}}
\tablewidth{0pt}
\tablehead{
\multicolumn{8}{c}{Theoretical Frequency\tablenotemark{a}} & \colhead{Mode Description\tablenotemark{b}} & \colhead{Mode No.\tablenotemark{b}} & \colhead{Observed Frequency\tablenotemark{c}} \\
\multicolumn{8}{c}{(cm$^{-1}$)} &  &  & \colhead{(cm$^{-1}$ / ($\mu$m))} \\
\colhead{431G} & \colhead{431Gd} & \colhead{631G} & \colhead{6311G} & \colhead{631Gd} & \colhead{631Gdp} & \colhead{631+Gdp} & \colhead{631++Gdp} &  &  & 
}
\startdata
3108.2 & 3127.3 & 3118.4 & 3076.2 & 3127.3 & 3128.8 & 3115.3 & 3083.1 &  $r_{\rm a}({\rm CH}_2)_{\rm v}$            & 78 &        \\ 
3113.6 & 3123.4 & 3113.6 & 3080.4 & 3122.7 & 3121.5 & 3119.5 & 3079.0 &  $r({\rm CH})$                              & 77 &        \\
3092.2 & 3103.1 & 3095.9 & 3072.6 & 3103.1 & 3102.2 & 3099.1 & 3071.8 &  $r({\rm CH})$                              & 76 & 3091.3 (3.23)\\
3077.2 & 3083.9 & 3076.5 & 3064.0 & 3083.8 & 3083.5 & 3081.2 & 3066.2 &  $r({\rm CH})$                              & 75 &        \\
3060.1 & 3066.9 & 3058.5 & 3062.9 & 3067.4 & 3067.3 & 3066.9 & 3065.7 &  $r({\rm CH})$                              & 74 & 3068.2 (3.26)\\
3052.7 & 3060.7 & 3054.6 & 3053.5 & 3060.6 & 3060.4 & 3057.4 & 3057.7 &  $r({\rm CH})$                              & 73 & 3061.4 (3.27)\\
3025.1 & 3051.3 & 3035.0 & 2999.1 & 3049.8 & 3045.9 & 3028.2 & 2996.6 &  $r({\rm CH}) + r({\rm CH}_2)_{\rm v}$      & 72 &       \\
3034.4 & 3045.2 & 3037.0 & 3044.2 & 3045.7 & 3045.5 & 3042.4 & 3047.2 &  $r({\rm CH}) + r({\rm CH}_2)_{\rm v}$      & 71 &       \\
3020.6 & 3033.4 & 3024.5 & 3051.4 & 3033.8 & 3033.9 & 3029.5 & 3053.9 &  $r({\rm CH})$                              & 70 & 3031.6\tablenotemark{d} (3.30)\\
3018.9 & 3028.2 & 3019.6 & 3039.6 & 3029.0 & 3029.2 & 3028.2 & 3045.8 &  $r({\rm CH})$                              & 69 &      \\
3015.4 & 3024.2 & 3015.7 & 3029.0 & 3025.3 & 3025.7 & 3023.9 & 3030.9 &  $r({\rm CH})$                              & 68 & 3019.0 (3.31)\\
2997.3 & 3000.5 & 2989.3 & 2986.0 & 3003.0 & 3002.8 & 3011.0 & 3007.3 &  $r({\rm CH})_{\rm v}$                      & 67 & 3011.0 (3.32)\\
1629.7 & 1656.4 & 1629.2 & 1619.5 & 1655.3 & 1654.8 & 1649.1 & 1647.1 &  $R($C$=$C$)_{\rm v} + \alpha({\rm HCH})_{\rm v} + \beta({\rm CCH})_{\rm v}$ &  66 &  1627.6\tablenotemark{d} (6.17)\\
1611.0 &  1634.0 &  1614.4 &  1608.6 &  1634.5 &  1634.8 &  1631.2 &  1632.3 &  $R({\rm CC}) + \beta({\rm CCH})$ &  65 &  1621.8\tablenotemark{d} (6.17)\\
1563.0 &  1581.0 &  1566.3 &  1561.7 &  1582.1 &  1582.3 &  1580.3 &  1580.2 &  $R({\rm CC}) + \beta({\rm CCH})$ &  63 &  1572.7\tablenotemark{d} (6.36)\\
1529.2 &  1551.6 &  1535.8 &  1529.9 &  1554.5 &  1555.6 &  1555.1 &  1560.4 &  $R({\rm CC}) + \beta({\rm CCH})$ &  62 &  1559.2\tablenotemark{d} (6.41)\\
1499.1 &  1518.9 &  1504.5 &  1498.4 &  1520.6 &  1521.0 &  1519.5 &  1523.3 &  $R({\rm CC}) + \beta({\rm CCH})$ &  61 &  1523.5 (6.56)\\
1481.5 &  1492.8 &  1483.0 &  1482.7 &  1493.3 &  1490.0 &  1490.1 &  1487.1 &  $R({\rm CC}) + \beta({\rm CCH})$ &  60 &  1487.8 (6.72)\\
1451.3 &  1457.4 &  1450.2 &  1450.6 &  1457.0 &  1453.6 &  1453.7 &  1447.6 &  $R({\rm CC}) + \beta({\rm CCH})$ &  59 &  1456.0 (6.87)\\
1442.0 &  1447.0 &  1441.5 &  1441.5 &  1446.3 &  1444.6 &  1443.7 &  1446.2 &  $R({\rm CC}) + \beta({\rm CCH})$ &  58 &  1445.4 (6.92)\\
1431.5 &  1430.5 &  1430.0 &  1431.0 &  1430.7 &  1425.4 &  1425.3 &  1425.5 &  $R({\rm CC})_{\rm v} + \alpha({\rm HCH})_{\rm v} + \beta({\rm CCH})_{\rm v}$ &  57 &  1429.0 (7.00)\\
1375.6 &  1391.5 &  1384.5 &  1375.5 &  1395.2 &  1395.7 &  1395.6 &  1400.3 &  $R({\rm CC}) + \beta({\rm CCH})$ &  56 &  1403.9 (7.12)\\
1397.6 &  1393.3 &  1394.7 &  1398.2 &  1391.6 &  1387.8 &  1389.5 &  1382.8 &  $R({\rm CC}) + \beta({\rm CCH})$ &  55 &  1389.5 (7.20)\\
1359.4 &  1365.7 &  1367.8 &  1359.3 &  1370.3 &  1369.6 &  1370.6 &  1376.1 &  $R({\rm CC}) + \beta({\rm CCH})$ &  54 &         \\
1320.2 &  1325.5 &  1324.1 &  1322.3 &  1327.4 &  1326.9 &  1327.4 &  1333.0 &  $R({\rm CC}) + R({\rm CC})_{\rm v} + \beta({\rm CCH}) + \beta({\rm CCH})_{\rm v}$ &  53 &  1342.2 (7.45)\\
1308.9 &  1315.5 &  1310.4 &  1309.3 &  1316.1 &  1315.8 &  1316.5 &  1318.6 &  $R({\rm CC}) + \beta({\rm CCH}) + \alpha({\rm HCH})_{\rm v}$ &  52 &  1315.2 (7.60)\\
1294.2 &  1293.3 &  1294.2 &  1296.0 &  1293.4 &  1288.3 &  1289.7 &  1289.8 &  $R({\rm CC}) + \beta({\rm CCH}) + \beta({\rm CCH})_{\rm v}$ &  51 &      \\
1283.9 &  1274.8 &  1281.9 &  1283.6 &  1273.7 &  1270.9 &  1273.6 &  1275.0 &  $R({\rm CC}) + \beta({\rm CCH}) + \beta({\rm CCH})_{\rm v}$ &  50 &  1280.5 (7.81)\\
1259.3 &  1261.5 &  1262.0 &  1256.0 &  1262.5 &  1261.2 &  1261.9 &  1255.6 &  $R({\rm CC}) + \beta({\rm CCH})$ &  49 &  1258.3 (7.95)\\
1233.4 &  1229.7 &  1235.0 &  1229.4 &  1230.7 &  1228.3 &  1230.1 &  1223.9 &  $R({\rm CC}) + \beta({\rm CCH}) + \alpha({\rm CCC})$  &  48 &  1223.6 (8.17)\\
1192.7 &  1189.4 &  1192.9 &  1191.6 &  1190.4 &  1187.9 &  1190.4 &  1182.0 &  $R({\rm CC}) + \beta({\rm CCH}) + \alpha({\rm CCC})$  &  47 &  1183.1 (8.45)\\
1188.8 &  1179.9 &  1186.5 &  1190.2 &  1178.7 &  1176.2 &  1178.7 &  1172.0 &  $\beta({\rm CCH})$ &  46 &  1177.3 (8.49)\\
1173.2 &  1164.9 &  1171.4 &  1167.6 &  1164.0 &  1162.0 &  1164.7 &  1154.3 &  $\beta({\rm CCH})$ &  45 &  1160.9 (8.61)\\
1153.6 &  1144.5 &  1153.7 &  1147.9 &  1144.4 &  1141.9 &  1141.6 &  1144.3 &  $\beta({\rm CCH})$ &  44 &  1142.6 (8.75)\\
1085.8 &  1080.1 &  1085.1 &  1086.4 &  1079.9 &  1077.3 &  1079.5 &  1078.8 &  $R({\rm CC})_{\rm v} + \beta({\rm HCH})_{\rm v} + \beta({\rm CCH})_{\rm v}$ &  42 &  1089.6 (9.18)\\
1019.0 &  1023.6 &  1020.5 &  1015.4 &  1023.5 &  1023.5 &  1022.9 &  1020.8 &  $R({\rm CC}) + \epsilon({\rm CH})_{\rm v} + \omega({\rm HCH})_{\rm v}$ & 41 & 1028.8 (9.72)\\
1008.6 &  1016.2 &  1009.8 &  1004.3 &  1016.2 &  1016.0 &  1016.2 &  1015.7 &  $R({\rm CC})$ &  40 &  1017.3 (9.83)\\
1004.4 &  1008.0 &  1003.8 &  1002.0 &  1006.7 &  1007.1 &  1004.8 &  995.7 &  $R({\rm CC}) + \epsilon({\rm CH})_{\rm v} + \omega({\rm HCH})_{\rm v}$ & 39 &  1000.9 (9.99)\\
978.1 &  976.2 &  977.4 &  981.0 &  975.0 &  975.3 &  976.9 &  979.1 &  $R({\rm CC})_{\rm v} + \beta({\rm HCH})_{\rm v} + \epsilon({\rm CH})_{\rm v} + \epsilon({\rm CH})$ &  38 &  992.2 (10.08)\\
986.2 &  963.6 &  981.7 &  988.6 &  961.0 &  965.7 &  972.3 &  971.2 &  $\epsilon({\rm CH})$ &  36 &  957.5 (10.44)\\
960.3 &  940.1 &  958.1 &  966.5 &  936.0 &  941.9 &  952.9 &  954.6 &  $\epsilon({\rm CH}) + \epsilon({\rm HCH})_{\rm v}$ &  35 &  953.6 (10.41)\\
955.8 &  937.6 &  953.2 &  962.1 &  933.2 &  939.2 &  950.1 &  949.1 &  $\epsilon({\rm CH}) + \epsilon({\rm HCH})_{\rm v}$ &  34 &  931.4 (10.74)\\
948.9 &  929.4 &  949.7 &  958.4 &  928.2 &  932.6 &  936.2 &  934.0 &  $\epsilon({\rm CH}) + \epsilon({\rm HCH})_{\rm v}$ &  33 &  926.6 (10.79)\\
918.8 &  908.4 &  915.1 &  918.2 &  906.0 &  906.8 &  909.5 &  914.8 &  $\alpha({\rm CCC}) + \epsilon({\rm CH})_{\rm v}$   &  32 &        \\
894.9 &  876.1 &  893.2 &  895.9 &  877.9 &  878.4 &  884.4 &  883.7 &  $\epsilon({\rm CH})$ &  31 &  883.2 (11.32)\\
855.2 &  847.7 &  856.3 &  856.9 &  849.0 &  850.3 &  849.7 &  849.3 &  $\epsilon({\rm CH})$ &  30 &  858.2 (11.65)\\
842.1\tablenotemark{e} &  844.4 &  843.5\tablenotemark{e} &  842.6\tablenotemark{e} &  845.5 &  846.5 &  846.2 &  847.4 &  $\alpha({\rm CCC}) + \epsilon({\rm CH})+ \beta({\rm CCH}) + \omega({\rm HCH})_{\rm v}$ & 29 &    \\ 
847.4 &  837.1 &  847.8 &  850.3 &  840.1 &  841.4 &  839.0 &  838.6 &  $\epsilon({\rm CH})$ &  28 &  842.7 (11.87)\\
843.8 &  840.0 &  841.4 &  840.3 &  837.3 &  838.2 &  841.1 &  844.5 &  $\alpha({\rm CCC})$  &  27 &        \\
788.2 &  780.6 &  788.6 &  788.5 &  783.9 &  782.7 &  779.2 &  783.0 &  $\epsilon({\rm CH})$ &  26 &  791.6 (12.63)\\
757.7 &  751.9 &  759.4 &  761.3 &  754.4 &  755.5 &  753.6 &  753.2 &  $\epsilon({\rm CH})$ &  25 &  755.0 (13.25)\\
739.2 &  733.0 &  736.7 &  735.2 &  735.4 &  735.8 &  730.7 &  734.3 &  $\epsilon({\rm CH}) + \tau({\rm CCC})$ &  24 &  740.5 (13.50)\\
737.0 &  731.3 &  738.3 &  738.7 &  733.4 &  734.2 &  732.0 &  731.0 &  $\epsilon({\rm CH}) + \tau({\rm CCC})$ &  23 &  734.7 (13.61)\\
685.4 &  683.8 &  687.7 &  689.8 &  684.6 &  685.3 &  684.5 &  685.5 &  $\omega({\rm HCH})_{\rm v}$ &  22 &  692.3 (14.44)\\
676.8 &  674.2 &  675.3 &  677.7 &  673.4 &  674.1 &  675.5 &  675.8 &  $\alpha({\rm CCC}) + \omega({\rm HCH})_{\rm v}$ &  21 &  680.8 (14.69)\\
636.4 &  629.1 &  634.6 &  636.4 &  627.9 &  628.6 &  630.1 &  637.1 &  $\alpha({\rm CCC})$ &  20 &  644.1 (15.53)\\
621.2 &  614.5 &  620.0 &  624.3 &  614.2 &  614.5 &  614.2 &  623.1 &  $\alpha({\rm CCC}) + \beta({\rm HCH})_{\rm v} + \tau({\rm CCC})$ &  19 &  627.7 (15.93)\\
594.2 &  585.8 &  592.2 &  595.6 &  584.7 &  585.4 &  586.1 &  591.5 &  $\alpha({\rm CCC}) + \omega({\rm HCH})_{\rm v}$ &  18 &  595.9 (16.78)\\
564.9 &  557.7 &  562.5 &  565.0 &  557.1 &  558.0 &  559.0 &  560.8 &  $\tau({\rm CCC})$ &  17 &  563.1 (17.76)\\
514.5 &  509.2 &  513.8 &  517.5 &  508.6 &  509.1 &  509.6 &  510.0 &  $\alpha({\rm CCC}) + \epsilon({\rm CH})_{\rm v} + \omega({\rm HCH})_{\rm v}$ &  16 &  513.9 (19.46)\\
500.2 &  498.5 &  501.4 &  500.0 &  499.6 &  499.6 &  499.7 &  498.4 &  $\tau({\rm CCC})$ &  15 &  504.3 (19.83)\\
418.1 &  416.5 &  417.8 &  419.3 &  416.5 &  417.1 &  416.9 &  418.3 &  $\tau({\rm CCC})$ &  12 &  420.4 (23.79)\\
400.7 &  395.2 &  399.2 &  402.0 &  394.6 &  395.3 &  396.9 &  403.2 &  $\alpha({\rm CCC}) + \tau({\rm CCC})_{\rm v}$ &  11 &  406.9 (24.58)\\
\enddata
\tablenotetext{a}{The frequencies have been scaled.}
\tablenotetext{b}{Derived from the results obtained with the 631Gdp basis set. See Table~\ref{tbl-2} for the notation of the modes.}
\tablenotetext{c}{Frequencies measured in Ar matrix unless indicated otherwise.}
\tablenotetext{d}{Band observed at room temperature in CsI pellet.}
\tablenotetext{e}{The atomic displacements ($\alpha({\rm CCC})+ \beta({\rm CCH})$) differ from those obtained with the other basis sets.}
\end{deluxetable}

\clearpage

\begin{deluxetable}{lcccc}
\tabletypesize{\scriptsize}
\tablecaption{Scaling Factors\label{tbl-4}}
\tablewidth{0pt}
\tablehead{
\colhead{Basis Set} & \multicolumn{2}{c}{2-VA} & \multicolumn{2}{c}{9-VA}\\
 & \colhead{Non-C$-$H Stretching} & \colhead{C$-$H Stretching} & \colhead{Non-C$-$H Stretching} & \colhead{C$-$H Stretching}
}
\startdata
4-31G	    &  0.9528 &  0.9519 &  0.9540 &  0.9567 \\
4-31Gd    &  0.9671 &  0.9582 &  0.9665 &  0.9619 \\
6-31G     &  0.9566 &  0.9534 &  0.9570 &  0.9546 \\
6-311G    &  0.9639 &  0.9582 &  0.9645 &  0.9619 \\
6-31Gd    &  0.9700 &  0.9533 &  0.9697 &  0.9603 \\
6-31Gdp   &  0.9726 &  0.9583 &  0.9715 &  0.9611 \\
6-31+Gdp  &  0.9769\tablenotemark{a} &  0.9584\tablenotemark{a} &  0.9756 &  0.9603 \\
6-31++Gdp &  0.9760\tablenotemark{b} &  0.9550\tablenotemark{b} &  0.9758 &  0.9575 \\
\enddata
\tablecomments{The scaling factors were applied to the raw theoretical harmonic frequencies to obtain the theoretical values given in Tables~\ref{tbl-2} and \ref{tbl-3}.}
\tablenotetext{a}{Geometry optimized using a tight grid.}
\tablenotetext{b}{Geometry optimized forcing a planar geometry.}
\end{deluxetable}

\clearpage

\begin{deluxetable}{lccccc}
\tabletypesize{\scriptsize} 
\tablecaption{Comparison of the Theoretical and Measured Spectra of 2-VA\label{tbl-5}}
\tablewidth{0pt}
\tablehead{
\colhead{Basis Set} & \multicolumn{2}{c}{Frequency Difference} & \colhead{Intensity Correlation} & \colhead{CPU Time} \\
 & \colhead{rms} & \colhead{Max. Dev.} & \colhead{$R^2$} & \\
 & \colhead{(cm$^{-1}$)} & \colhead{(cm$^{-1}$)} &  & \colhead{(min)}
}
\startdata
4-31G	     &  12.8 &  42.2 &  0.8730 &  117.6 \\
4-31Gd     &  13.2 &  40.8 &    0.8951 &  253.0 \\
6-31G      &  11.2 &  38.6 &    0.8961 &  148.8 \\
6-311G     &  13.7 &  45.9 &    0.9035 &  221.4 \\
6-31Gd     &  13.2 &  37.2 &    0.8957 &  225.9 \\
6-31Gdp    &  13.0 &  38.6 &    0.8890 &  242.8 \\
6-31G+dp\tablenotemark{a}  &   9.5 &  33.9 & 0.7384 &  3994.2 \\
6-31G++dp\tablenotemark{b} &  11.7 &  36.2 & 0.8517 &  816.0  \\
\enddata
\tablecomments{Only the bands with a frequency between 400--1800 cm$^{-1}$ have been taken into account.}
\tablenotetext{a}{Geometry optimized using a tight grid.}
\tablenotetext{b}{Geometry optimized forcing a planar geometry.}
\end{deluxetable}

\clearpage

\begin{deluxetable}{lccccc}
\tabletypesize{\scriptsize} 
\tablecaption{Comparison of the Theoretical and Measured Spectra of 9-VA\label{tbl-6}}
\tablewidth{0pt}
\tablehead{
\colhead{Basis Set} & \multicolumn{2}{c}{Frequency Difference} & \colhead{Intensity Correlation} & \colhead{CPU Time} \\
 & \colhead{rms} & \colhead{Max. Dev.} & \colhead{$R^2$} & \\
 & \colhead{(cm$^{-1}$)} & \colhead{(cm$^{-1}$)} &  & \colhead{(min)}
}
\startdata
4-31G	     & 11.9 &  30.0 &  0.8180 &  130.1 \\
4-31Gd     &  9.0 &  28.8 &    0.6840 &  297.1 \\
6-31G      & 10.3 &  24.3 &    0.7335 &  174.4 \\
6-311G     & 12.9 &  31.8 &    0.8627 &  262.8 \\
6-31Gd     &  8.7 &  27.7 &    0.7149 &  271.9 \\
6-31Gdp    &  8.5 &  27.2 &    0.4092 &  264.3 \\
6-31G+dp   &  8.4 &  21.5 &    0.8172 &  689.0 \\
6-31G++dp  &  7.1 &  19.5 &    0.8663 &  1778.0 \\
\enddata
\tablecomments{Only the bands with a frequency between 400--1800 cm$^{-1}$ have been taken into account.}
\end{deluxetable}

\end{document}